\documentclass[aps,pra,reprint,showpacs,superscriptaddress]{revtex4-1}
\usepackage[english]{babel}
\usepackage{amsmath,amssymb,bbm,graphicx,color,comment,txfonts}
\usepackage[bookmarks=true,colorlinks,citecolor=blue,urlcolor=blue]{hyperref}
\usepackage{dsfont}
\usepackage[normalem]{ulem}
\usepackage{float} 
\usepackage[ruled,vlined]{algorithm2e}

\newcommand{\titleofthepaper}{A learning algorithm with emergent scaling behavior for classifying phase transitions}

\begin{document}

\title{\titleofthepaper}

\author{N. Maskara}
\email{nmaskara@g.harvard.edu}
\affiliation{%
 Division of Physics, Mathematics and Astronomy, California Institute of Technology, Pasadena, CA 91125, USA
}
\affiliation{%
 Department of Physics, Harvard University, Cambridge, MA 02138, USA
}
\author{M. Buchhold}%
\affiliation{%
 Division of Physics, Mathematics and Astronomy, California Institute of Technology, Pasadena, CA 91125, USA
}%
\affiliation{%
 Institute for Theoretical Physics, University of Cologne, Z\"ulpicher Strasse 77, 50937, Cologne, Germany
}%

\author{M. Endres}
\affiliation{%
 Division of Physics, Mathematics and Astronomy, California Institute of Technology, Pasadena, CA 91125, USA
}%
\author{E. van Nieuwenburg}%
\affiliation{%
 Division of Physics, Mathematics and Astronomy, California Institute of Technology, Pasadena, CA 91125, USA
}%
\affiliation{%
 Niels Bohr International Academy, Niels Bohr Institute, University of Copenhagen, Blegdamsvej 17, 2100 Copenhagen, Denmark
}%
\date{\today}

\begin{abstract}
Machine learning-inspired techniques have emerged as a new paradigm for analysis of phase transitions in quantum matter.
In this work, we introduce a supervised learning algorithm for studying critical phenomena from measurement data, which is based on iteratively training convolutional networks of increasing complexity, and test it on the transverse field Ising chain and $q=6$ Potts model.
At the continuous Ising transition, we identify scaling behavior in the classification accuracy, from which we infer a characteristic {\it classification} length scale. 
It displays a power-law divergence at the critical point, with a scaling exponent that matches with the diverging correlation length.
Our algorithm correctly identifies the thermodynamic phase of the system and extracts scaling behavior from projective measurements, independently of the basis in which the measurements are performed.
Furthermore, we show the classification length scale is absent for the $q=6$ Potts model, which has a first order transition and thus lacks a divergent correlation length.
The main intuition underlying our finding is that, for measurement patches of sizes smaller than the correlation length, the system appears to be at the critical point, and therefore the algorithm cannot identify the phase from which the data was drawn.
\end{abstract}

\maketitle

{\it Introduction.}~--
Machine learning techniques have emerged as a new tool for analyzing complex many-body systems~\cite{carrasquilla_machine_2020, mehta_high-bias_2019}.
A particularly well-studied application of such techniques is that of the identification and classification of phase transitions directly from data, assuming little to no prior knowledge of the underlying physics~\cite{van_nieuwenburg_learning_2017, carrasquilla_machine_2017, wang_discovering_2016, giannetti_machine_2019, ponte_kernel_2017, beach_machine_2018, liu_learning_2019, wetzel_machine_2017, schafer_divergence_2019, Mendes_Santos_2021}.
Recent efforts have expanded such explorations to a diverse range of systems including disordered~\cite{venderley_machine_2018, theveniaut_neural_2019, van_nieuwenburg_learning_2018} and topologically ordered systems~\cite{zhang_interpretable_2019, deng_machine_2017, rodriguez-nieva_identifying_2019, balabanov_unsupervised_2020, greplova_unsupervised_2019}, as well as applications to experiments~\cite{rem_identifying_2019,bohrdt_classifying_2019,Zhang_2019_electronic}. 

An often-voiced concern, however, is that machine learning methods appear as a black box and that it is difficult to trust neural network classification without traditional supporting evidence.
For example, in the study of phase transitions, the phase boundary identified by a machine learning algorithm may be affected by short-distance correlations, which turn out to be irrelevant for the thermodynamic phase of the system~\cite{beach_machine_2018}.
Instead, learning algorithms should ideally focus on phase transition features which characterize the transition, such as power-law divergences near the critical point of a second order phase transition.

In this paper, we develop a machine learning algorithm inspired by this fundamental feature of critical phenomena, i.e. the emergence of long-distance correlations and scale invariance. 
The algorithm systematically analyzes critical behavior near a suspected transition point, using only snapshot data, by varying the functional form of a neural network. 
Specifically, we restrict the architecture so the network can only access patches of the snapshot at a time, and then vary the largest patch size. The resulting architecture is similar to those in Refs~\cite{wetzel_machine_2017, szegedy_going_2014, lin_network_2014, mills_extensive_2019, miles_correlator_2020}. 
We observe that, under these conditions, we can extract information about the spatial growth of correlations in the underlying data from the behavior of the classification probabilities.

\begin{figure}[t]
    \centering
    \includegraphics[width=0.45\textwidth]{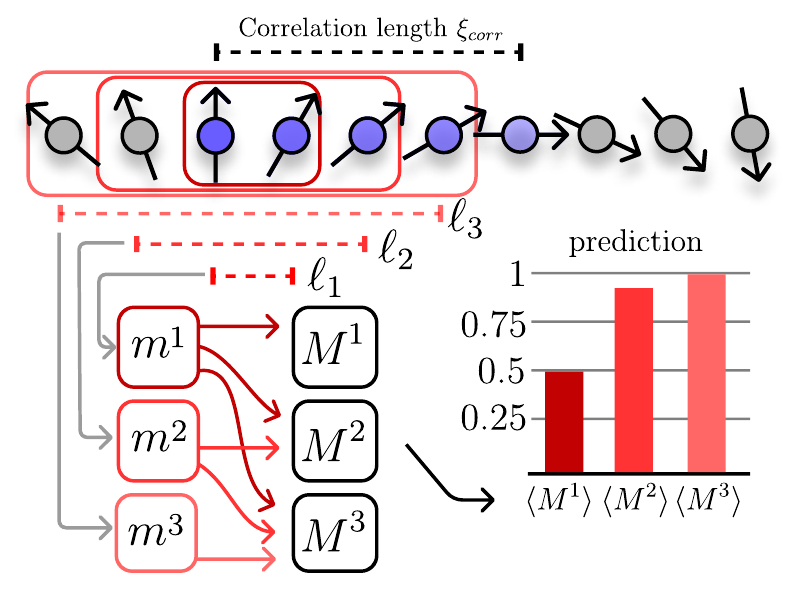}
    \caption{Conceptual illustration of our method for a 1D spin-chain. Snapshots near a second order phase transition reveal a characteristic length scale $\xi_{\text{corr}}$, over which spins are correlated, and which diverges at the critical point. 
    The modules $m^k$ are designed to only capture correlations in the data up to a certain length scale $\ell_k$, and their outputs are aggregated into $M^{k}$.
    On length scales shorter than $\xi_{\text{corr}}$, the algorithm cannot make a firm distinction between the two phases. As the module size is increased, the prediction $\langle M^{k} \rangle$ is improved until $\ell_k \sim \xi_{\text{corr}}$, after which it quickly saturates.}
    \label{fig:concept}
\end{figure}

Our main result is the identification of an emergent length scale, which we extract from classification data, and which displays scaling behavior. 
Physical arguments suggest that this {\it classification} length scale reflects the system's correlation length, which diverges at the critical point according to a power law with a universal exponent.
We exemplify this on the one-dimensional (1D) transverse field Ising model, whose critical exponents are known, and compare the scaling of the physical correlation length with the classification length.
We also consider a first order transition in the two-dimensional (2D) $q=6$ Potts model, where, in-line with expectations, no scaling of the classification length is observed.

The learning algorithm proves quite versatile, and neither requires prior knowledge of the structure of the order parameter nor of the measurement data. We demonstrate this by considering projective measurements in different measurement bases, with which the thermodynamic phase and its order parameter cannot be readily inferred from conventional two-point correlation functions.
Nevertheless, the algorithm is capable of learning more complex correlations, and manages to distinguish the two phases with a high degree of accuracy. 
This is especially promising in light of studying phase transitions with non-local or hidden order, for which algorithms based on purely local structures hardly gain access.

{\it The Algorithm.}~-- We start by introducing a supervised learning algorithm that allows one to systematically add complexity to a machine learning model. 
The model is composed of a set of independent computational units, termed {\it modules}.
The algorithm takes and trains modules iteratively, and, by design, each new module learns correlations in the data that the prior modules did not capture.
Conceptually, complexity is added by increasing the amount of correlations representable by the model in each step. 
Here, we are interested in scaling behaviour near critical points, so each subsequent module is designed to capture spatial correlations at a larger length-scale.

Each module $m^i: \vec{x} \rightarrow \mathbb{R}$, labeled with an index $i$, takes as input a snapshot (projective measurement) $\vec{x}$ and maps it to a scalar.
Then, we apply an aggregation function $M^k$ that aggregates the outputs of the first $k$ modules $m^1,...,m^k$.
In practice, both the modules and the the aggregation function are implemented through a neural network. 
The modules plus the aggregation function constitute our combined machine learning model $M^k(m^1,...,m^k): \vec{x} \rightarrow \mathbb{R}$, mapping an input $\vec{x}$ to a corresponding classification target $y$.
This model is then trained to minimize a classification loss function $\mathcal{L}$ using the iterative training algorithm described in pseudo-code in \textbf{Algorithm 1}:

\vspace{0.2cm}
\begin{algorithm}[H]
\caption{Iterative Training Algorithm}
\SetKwInOut{Input}{input}\SetKwInOut{Output}{output}
\Input{Sequence of modules $m^1,...,m^l$}
\Input{Aggregate models $M^k(m^1,...,m^k), k\le l$}
\Input{Labeled dataset $\{(\vec{x}_i,y_i)\}$}
\Input{Loss function $\mathcal{L}$}
\KwResult{Trained set of models $\{M^k\},k=1,...,l$}
\For{$k=1,...,l$}{
train $M^k$ on dataset $(\vec{x}_i,y_i)$\ by minimizing $\mathcal{L}(y_i, M^k(x_i))$; \\
freeze parameters of $m^k$\;
}
\end{algorithm}
\vspace{0.2cm}

To understand the intuition behind this algorithm, it is helpful to consider an aggregation function which simply sums the module outputs, $M^k(\vec{x}) = \sum_{j=1}^k m^k(\vec{x})$, and a loss function like mean-squared error, which only depends on the difference between the target dataset and model output.
In this case, the loss function can be written as
\begin{align}
    \mathcal{L}\left(y_i - M^k(\vec{x}_i)\right) = \mathcal{L}\left(\tilde{y}_i -  m^k(\vec{x}_i)\right),
\label{eq:loss}
\end{align}
where we define an effective target $\tilde{y}_i = y_i - \sum_{j=1}^{k-1} m^j(\vec{x}_i)$.
Crucially, during the $k$-th step of training, the variational parameters (weights) of all the modules $m^j,j<k$ are frozen, and hence $m^k$ is trained only on the \textit{residuals} $\tilde{y}_i$ of all the prior models. 
These residuals are the leftover errors from the previous training step, and each subsequent module only learns features which prior modules did not capture.
In our actual implementation, we use a linear classifier for the aggregation function and the binary cross-entropy for the loss function (see below). As a result, our modules are not exclusively trained on residuals, but the intuitive picture of subtracting prior features from the cost function still approximately holds.

\begin{figure*}[t]
    \centering
    \includegraphics[width=0.98\textwidth]{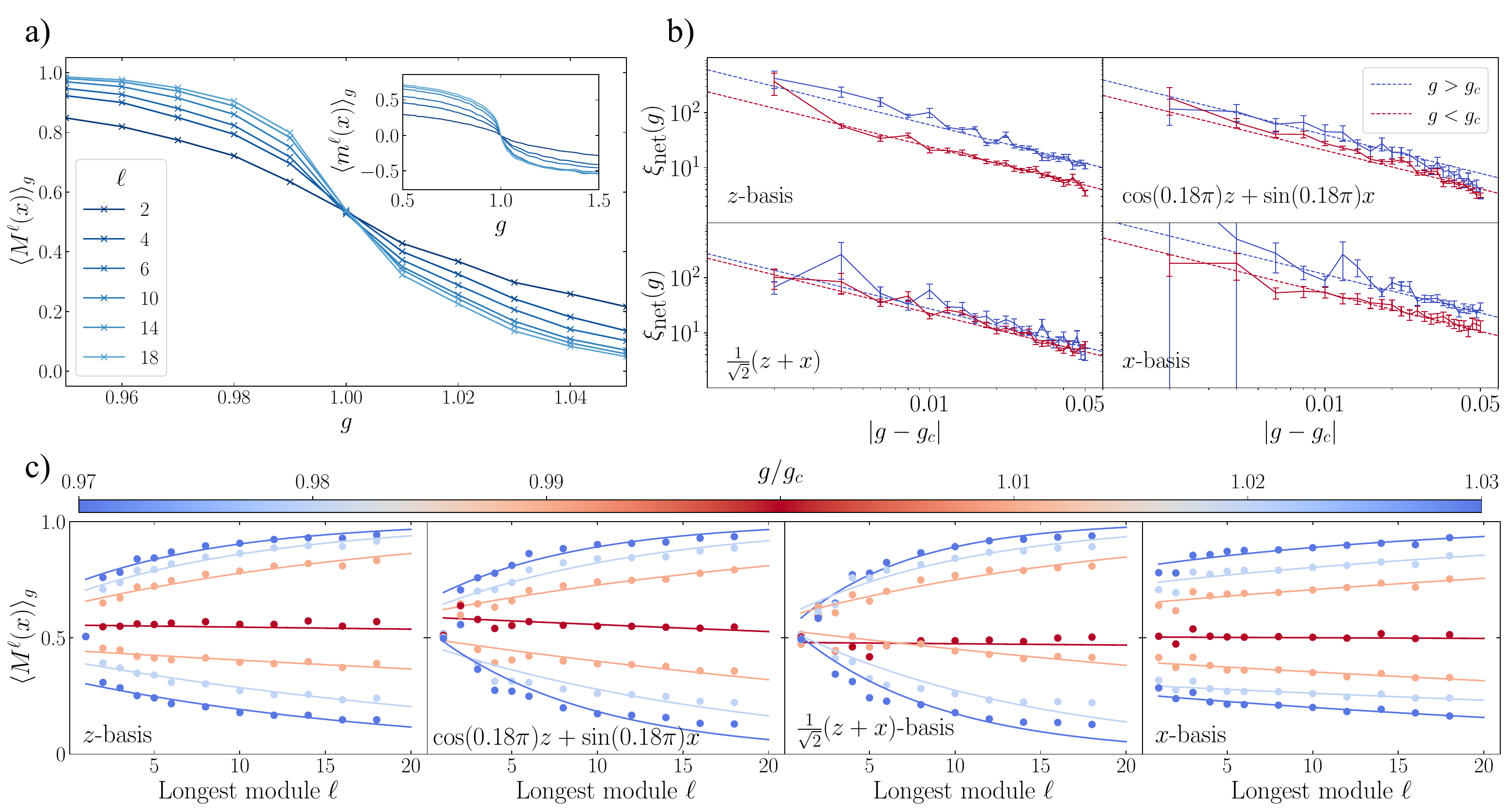}
    \caption{
    (a) Average classification $M^\ell(x)$ as a function of the TFIM parameter $g$ and the spatial extent $\ell$ of the model. Classifiers incorporating longer range correlations can more reliably identify the phase of snapshots taken near the critical point, as evidenced by an increase in the slope.
    (inset) Average module outputs $\langle m^\ell\rangle$ also exhibit scaling with $\ell$, as one would expect from an order parameter. 
    (b) Fitted classification lengths $\xi_{\text{net}}$ on a log-log plot, from measurements in the $z$ (basis $=0$), $x$ (basis $=\pi/2$), and two intermediate bases $\cos(0.18\pi)z + \sin(0.18\pi)x$ (basis $=0.18\pi)$ and $(z+x)/\sqrt{2}$ (basis $=\pi/4$), plotted on a log-log scale. For each basis, points are separate for approaching $g_c$ from above (blue) or below (red). Data for $g$ near $g_c$ is consistent with power-law behavior with exponent $\mu=1$, which are depicted by colored dotted lines.
    Error bars are an estimate of the standard deviation in the fitted $\xi_{\text{net}}$.
    (c) Correlation lengths $\xi_{\text{net}}$ are extracted by fitting the classification curves $\langle M^\ell\rangle_g$ for each value of $g$, to an inverse exponential form. Fits are performed for classifiers with largest module size between $2 \leq \ell < 20$, and are shown in the same four bases. Notice in the intermediate bases, especially $\pi/4$, the classification curves behave erratically at very small $\ell$, but exhibit scaling at intermediate $\ell > 5$. The resulting lengthscales extracted by the exponential fits are consistent with universal scaling near the critical point $g \approx g_c$ (b).
    }
    \label{fig:classification_scaling}
\end{figure*}

{\it The Network.}~-- In this section, we'll provide more details about our choice of neural networks. 
For the modules, we develop a class of convolutional neural networks designed to probe the spatial locality of correlations.
Each module $m^k$ takes the functional form
\begin{align}
    m^k(\vec{x}) &= \frac{1}{N} \sum_j  m^{k}_0(\vec{x}_j^{\ell_k})
\end{align}
where $m^{k}_0$ is a two layer neural network that acts on a subset  $\vec{x}_j^{\ell_k}$ of the data $\vec{x}$. 
The label $\ell_k$ indicates the size of a spatial region
corresponding to the subset (e.g., $\ell_k$ adjacent sites in a one-dimensional lattice) and the index $j$ enumerates all $N$ regions of size $\ell_k$. 

The aggregation function we choose is a linear classifier $LC$, acting on the module outputs $m^k(\vec{x})$.
\begin{align}
    M^{K}(\vec{x}) &= LC\Big(m^{1}(\vec{x}), m^{2}(\vec{x}),...m^{K}(\vec{x}) \Big).
\end{align}
The linear classifier is defined by $LC(\{m^i(\vec{x})\}) = \sigma(\sum_i w_i m^i(\vec{x}) - b)$, where $\sigma$ is the sigmoid (or logistic) function $\sigma(z)=1/(1+e^{-z})$, and $w_i$ and $b$ are free parameters.
The non-linearity $\sigma$ maps the linear combination of module outputs to a value between $[0,1]$, as expected for binary classification.
The loss function we use is the binary cross entropy, $\mathcal{L}(y_i, M^k(x_i)) = \langle y \log M^k(x_i) + (1-y) \log (1-M^k(x_i)) \rangle$, where $y_i$ is the target label $y_i \in \{0,1\}$ and the expectation value is taken over the training dataset.

Choosing a linear classifier ensures that the network cannot capture additional, spurious correlations between modules. 
For example, for a linear classifier, the first module that extracts information on a $5$-site region will be the module $m^5(\vec{x})$. 
Instead, a non-linear classifier, which contained e.g. terms quadratic in the arguments, would also include products of the form $m^2(\vec{x})m^3(\vec{x})$. This would include \textit{non-local} information about potentially disconnected $5$-site regions, which we exclude on purpose. 
Thus, in the remainder of the text, we use $M^\ell$ instead of $M^K$ to denote an aggregate classifier with largest convolution length $\ell_K = \ell$.

The full model is naturally represented by a convolutional neural network (CNN), which can be implemented using open source machine learning libraries such as Keras~\cite{chollet2015keras} or PyTorch~\cite{NEURIPS2019_9015}. 
The resulting CNN architecture, with different parallel convolutional modules, is similar to the correlation probing neural network from Ref.~\cite{miles_correlator_2020}, the network-in-network architectures from Refs.~\cite{wetzel_machine_2017, szegedy_going_2014, lin_network_2014}, and to the \textit{EDNN} from Ref.~\cite{mills_extensive_2019}. 
However, to the best of our knowledge, scaling the convolution size by freezing the module parameters iteratively, and observing the response via classification accuracy to extract a lengthscale, has not appeared before in the literature.

\vspace{0.2cm}
{\it Applications.}~--
To investigate the scaling of the classification output of our model, we first analyze a second order phase transition in the paradigmatic 1D transverse field Ising model (TFIM). 
We then contrast this behaviour with a first order transition in the 2D $q=6$ Potts model.

The single-parameter Hamiltonian for the 1D TFIM with open boundaries is
\begin{align}
    H(g) = -\sum_{i=1}^{L-1} \sigma_{i}^z \sigma_{i+1}^z - g\sum_{i=1}^{N} \sigma_i^x, \label{eq:TFIM}
\end{align}
where $\sigma^{z,x}_i$ are Pauli matrices for spin $i$.
At critical value $g_c=1$, the ground state of this model undergoes a phase transition from a disordered (paramagnetic) to an ordered (ferromagnetic) state, breaking the global $\mathds{Z}_2$-symmetry.
In what follows we focus our attention on a region around $g=g_c$.
To construct our dataset, we employ the matrix product state based iTEBD algorithm~\cite{vidal_classical_2007} to numerically determine the ground state as a function of $g$, and sample configurations for a system size of $L = 400$.
We then perform projective measurements in multiple bases, including the $z$-basis (measuring $\sigma^z_i$ on each site), but also in the $x$-basis (measuring $\sigma^x_i$) and in a few intermediate basis, $\cos(\theta) \sigma^z_i + \sin(\theta) \sigma^x_i$. 
This is done to illustrate that the classification algorithm does not rely on the a priori choice of an optimal basis, which for experimental measurements may be unknown.

Each snapshot is labelled with the phase it is drawn from, e.g. ordered ($g < g_c$) or disordered ($g > g_c$). 
As a convention we set the label equal to $1$ if the snapshot $x$ is drawn from the ordered phase.
The machine learning model is then trained by minimizing the binary cross-entropy between the labels and the prediction $M^\ell(x)$ (see \textit{The Network}), on snapshots drawn from the ground state of $H(g)$ at 85 different values of $g$ with 800 snapshots per $g$. Points were spread from $g=0$ to $4.4$, but concentrated in the critical region near $g_c=1$, with a minimum separation of $\Delta g = 0.01$.
Module outputs, $\langle m^\ell \rangle_g$, and phase classification $\langle M^\ell \rangle_g$, are computed on a separate validation dataset consisting of $200$ snapshots per $g$, and with a minimum separation of $\Delta g = 0.002$ in the critical region.
The value of $\langle M^\ell \rangle_g$ is a measure of how accurately one can identify the phase of the ground state of $H(g)$ from local measurements with maximum spatial extent $\ell$, and henceforth we call this the classification accuracy. 

Empirically, we find the classification accuracy improves as the convolution size $\ell$ is increased, with the improvement most dramatic for snapshots drawn near $g_c$, as shown in Fig.~\ref{fig:classification_scaling}a.
To rationalize this behavior, we notice that the problem of identifying the phase from local measurements is intimately connected to the correlations in the ground state. 
At the critical point, the ordered and disordered phases are indistinguishable, and the system is dominated by fluctuations with a divergent correlation length $\xi_{\text{corr}}$.
This results in an ambiguous prediction $\langle M^\ell \rangle_{g_c}=0.5$. 
As $g$ moves away from the critical point, characteristic correlations, indicating one of the possible phases, start to build up at distances larger than $\xi_{\text{corr}}$, while shorter distances remain dominated by fluctuations. 
Since $\xi_{\text{corr}}\sim \vert g-g_c \vert^{-\nu}$ follows a power-law, farther from the critical point the classification accuracy improves for fixed $\ell$. Similarly, as $\ell$ is increased, the accuracy of the classification $\langle M^\ell \rangle_g$ will improve, saturating to either 1 or 0  depending on $g < g_c$ or $g > g_c$ respectively. 

These arguments suggest the behavior of the classification curves, $\langle M^\ell \rangle_{g}$ can be used to detect a second order phase transition.
Indeed, the improvement of $M^\ell$ with $\ell$ is contingent upon correctly partitioning the dataset close to the critical point, and is unique to the critical region. Away from the critical point, the  classification probabilities saturate at some constant $\ell$ and no visible improvement for larger $\ell$ is observed~\cite{SOM}.

Associated with the behavior of $\langle M^\ell \rangle_{g}$ is a characteristic length scale, which we extract via an exponential fit $\langle M^{\infty}-M^\ell \rangle_g \sim \exp(-l/\xi_{\text{net}})$, where $\xi_{\text{net}}$ is a $g$-dependent length scale.
Specifically, we set $M^{\infty}$ to $1$ in the ordered phase and $0$ in the disordered phase, the saturation values in an ideal, infinitely large system. 
It turns out to be sufficient to consider $\ell < 20$ in order to obtain convincing prediction probabilities for datapoints with $|g-g_c|\geq 0.01$. For our data, which is for an $L=400$ chain, this then emulates a thermodynamically large system.
We also exclude the smallest module ($\ell=1$), which only captures single-site observables.
The resulting fit between $2 \leq \ell < 20$ performs well in the $z$-basis (~\ref{fig:classification_scaling}c). However, in different bases, specifically the intermediate $\theta=\pi/4$ basis, the classification accuracy exhibits erratic behavior for small $\ell$. Nevertheless, at slightly larger $\ell\approx 5$, scaling with $\ell$ reappears, and the exponential fit extracts a meaningful length scale $\xi_{\text{net}}$ capturing the scaling of $\langle M^\ell \rangle_{g}$.

Near the phase transition, the fitted correlation length $\xi_{\text{net}}$ diverges as $g$ approaches the critical point $g_c$, and is well described by a power-law (Fig. \ref{fig:classification_scaling}b).
It is known that, for the 1D TFIM, the physical correlation length scales as $\xi_{\text{corr}}$\,$\sim$\,$|g-g_c|
^{-\nu}$ with $\nu=1$. 
Remarkably, the power-law scaling of the fitted classification length $\xi_{\text{net}}\sim |g-g_c|^{-\nu_{\text{net}}}$, is consistent with $\nu_{\text{net}} \approx\nu= 1$ in any of the measured bases. This also includes the intermediate $\theta = \pi/4$ basis, where the phases cannot be reliably distinguished from conventional two-point or string-like correlation functions~\cite{SOM}. 
These observations suggests that the characteristic lengthscale we extracted by scaling the machine learning model reflects the underlying correlation lengthscale, and hence can be used to probe the growth of correlations near the critical point.

\begin{figure}[t]
    \centering
    \includegraphics[width=0.48\textwidth]{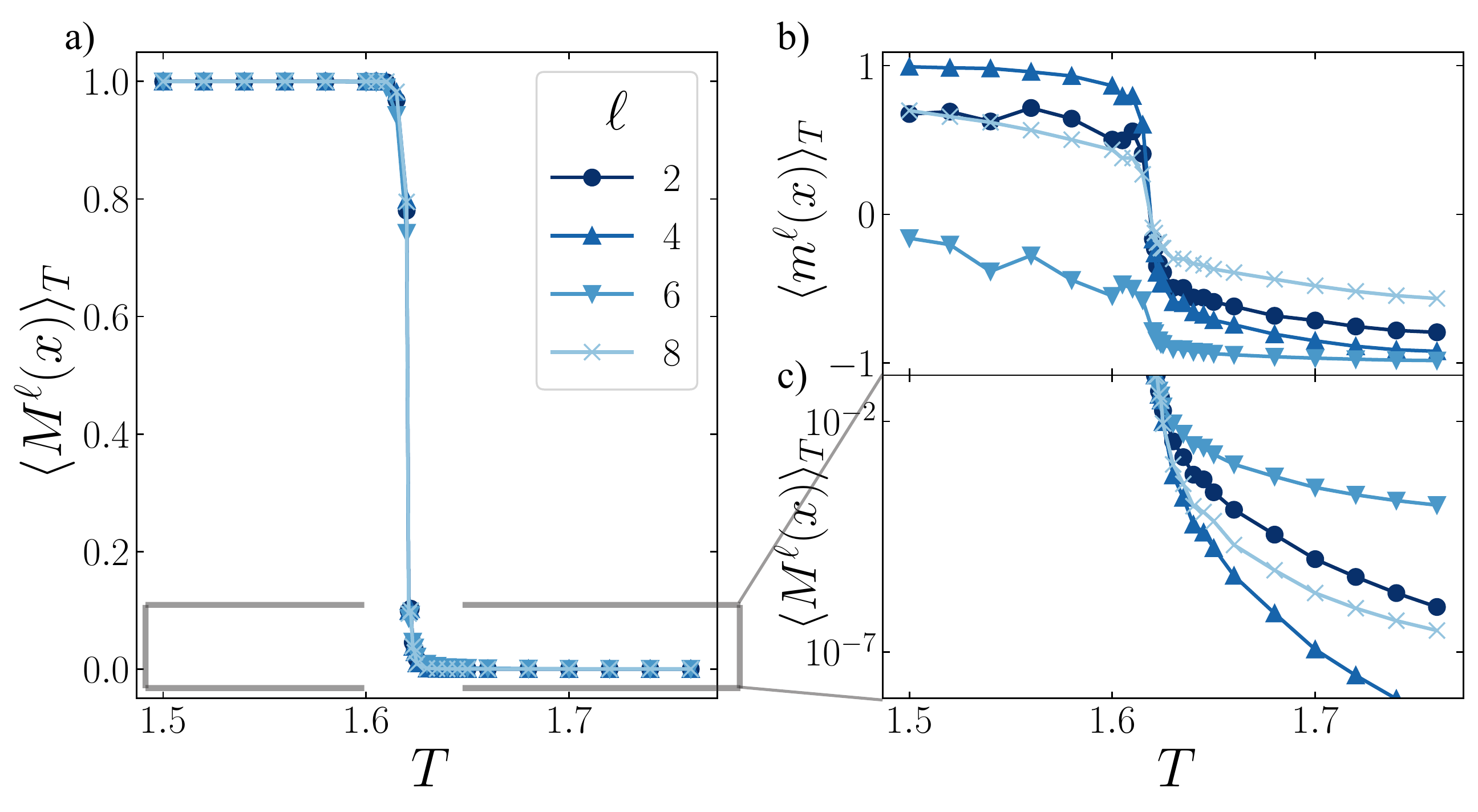}
    \caption{ Potts Model Data. 
    (a) Predicted classifications $\langle M^\ell \rangle_T$ for the $q=6$ Potts model do not exhibit scaling with system size, reflecting the fact that the phase transition is first order.
    (b) The module outputs $\langle m^\ell \rangle_T$, averaged over all snapshots at a given temperature, are in sharp contrast to the second order transition from Fig.~\ref{fig:classification_scaling}.
    (c) Zooming in on the classifications $\langle M^\ell \rangle_T$ near zero demonstrates the lack of scaling with $\ell$.
    }
    \label{fig:potts_model_data}
\end{figure}

To contrast this, we also examine the $q=6$ Potts model, which exhibits a first-order phase transition in temperature ($T$), and hence does not feature a diverging correlation length at the critical point~\cite{PottsModel}. 
As a result, we expect that spatially local measurement data should be sufficient to distinguish the two phases, even arbitrarily close to the transition point, since there are no long-range critical fluctuations.
Indeed, our numerical data reflects this intuition, since the classification $\langle M^\ell \rangle_{T}$ do not exhibit improvement beyond $\ell=2$ (Fig.~\ref{fig:potts_model_data}), regardless of distance to the transition point. 

{\it Discussion.}~-- We have presented a flexible method for extracting a characteristic length scale $\xi_{\text{net}}$ inferred from measurement data via scaling of the classification $\langle M^\ell \rangle_{g}$. 
Furthermore, $\xi_{\text{net}}$ diverges with the same scaling exponent as the physical correlation length $\xi_{\text{corr}}$, suggesting the two quantities coincide near the critical point.
Above we gave physical arguments for why this is the case, based on indistinguishability of the phase at short distances dominated by critical fluctuations.
These arguments are quite general, and reflect the emergence of universality at critical points.
Indeed, it is widely believed that at the critical point, the correlation length
$\xi_{\text{corr}}$ is the sole contributor to singular, thermodynamic quantities~\cite{kardar_statistical_2007, sachdev2011}. 
As long as the model and training data are sufficient to distinguish the two phases, the trained phase classification $\langle M^\ell \rangle_{g}$ should exhibit a discontinuitity at the critical point $g=g_c$ in the thermodynamic limit $\ell \rightarrow \infty$, and $\xi_{\text{corr}}$ should be the only length scale governing long-distance behavior of $\langle M^\ell \rangle_{g}$, in agreement with our observations.
As such, our scaling procedure makes few assumptions about the microscopic model, and could be used to study any system with a continuous phase transition.

Our analysis of the classification length scale $\xi_{\text{net}}$, has been performed for relatively small module sizes $\ell < 20$ compared to the size of the system $L=400$. This mitigates any potential finite size effects and yields a scaling of the classification only with the correlation length. 
In general, however, when the system size becomes comparable to the correlation length $\xi_{\text{corr}}\sim L$, both length scales will be relevant for characterizing the classification curves $\langle M^\ell \rangle_{g}$.
One could explore scaling with both parameters, to extend our framework to finite size systems.
Such scenarios will likely be relevant for near-term quantum simulators, where resources are limited.

The framework presented here for analyzing continuous phase transitions shows more broadly that for classification tasks, we can probe an underlying feature of interest by systematically varying the functional form of the models, and measuring the response in the form of classification accuracy. 
In addition to the convolutional neural networks used here, one could consider different classes of models, including kernel methods~\cite{giannetti_machine_2019,liu_revealing_2020}, or quantum machine learning models~\cite{Cong_2019}. 
Such a concept could also be readily generalized to different kinds of order.
For example, dynamical critical exponents could be estimated by looking at time-series data, and using a set of models which can represent observables spanning a finite-number of time-slices~\cite{van_nieuwenburg_learning_2018}. 
Similarly, phase transitions without conventional order parameters could be studied, perhaps by looking at loop-like observables of different sizes as in Ref.~\cite{zhang_quantum_2017}.
The tools developed here are especially timely, as quantum simulation technologies are rapidly approaching the point where exotic phases of quantum matter can be directly studied in the lab. 
The ability of our algorithm to characterize phase transitions and implicitly identify useful correlations, directly from measurement data, makes it a promising tool for analysis of future experiments.

\begin{acknowledgments}
We thank Dolev Bluvstein, Dan Borgnia, Iris Cong, Brian Timar, and Ruben Verresen for insightful discussions.
N.M. acknowledges funding from the Department of Energy Computational Science Graduate Fellowship under Award Number(s) DE-SC0021110. M.B. acknowledges funding via grant DI 1745/2-1 under DFG SPP 1929 GiRyd.
This project has received funding from the European Union’s Horizon 2020 research and innovation program under the Marie Sklodowska-Curie grant agreement No. 847523 ‘INTERACTIONS’, and the Marie Sklodowksa-Curie grant agreement No. 895439 `ConQuER'.
\end{acknowledgments}


\bibliography{references}

\clearpage

\onecolumngrid

\begin{center}
	\textbf{\large Supplementary material for ``\titleofthepaper'' }\\[5pt]
\end{center}
\setcounter{equation}{0}
\setcounter{figure}{0}
\setcounter{table}{0}
\setcounter{page}{1}
\setcounter{section}{0}
\makeatletter
\renewcommand{\theequation}{S\arabic{equation}}
\renewcommand{\thefigure}{S\arabic{figure}}
\renewcommand{\thepage}{S\arabic{page}}

\twocolumngrid

\section{Network architecture}

\begin{figure*}
    \centering
    \includegraphics[width=0.95\textwidth]{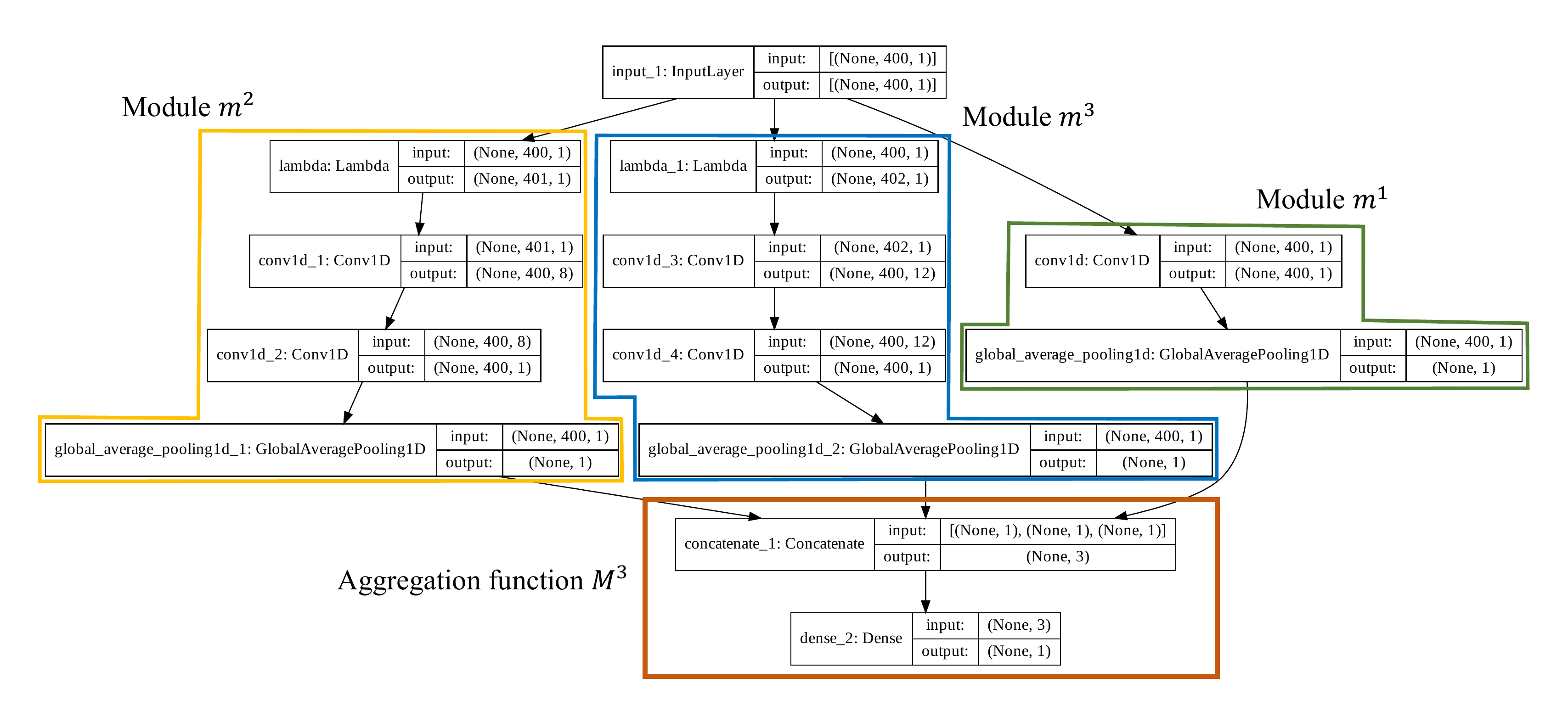}
    \caption{Network architecture at the third iteration, calculating the function denoted $M^3$. The three modules $m^1,m^2,m^3$ are convolutional layers, added in parallel, and are combined by a standard linear classifier. The Lambda layer is a custom layer, added to simulate convolutions which wrap around the periodic boundaries.
    The activation function for the first convolution (conv1d\_1,conv1d\_3) is a ReLU, while the activation for the second convolution (conv1d\_2,conv1d\_4) is tanh. The final dense layer has a sigmoid activation. Higher iterations have more modules added in parallel.
    For the Potts model, an equivalent architecture but with 2D convolutions was used. }
    \label{fig:example_network}
\end{figure*}

In this section, we provide a few additional details on our neural network implementation. As explained in the main text, each module $m^k$ is a two layer convolutional neural network. The first convolution takes an $\ell_k$ sized input to a vector of $4\ell_k$ features, and applies a ReLU non-linearity defined by $\text{ReLU}(x)=\max(0,x)$. Then, the second convolution collapses the $4\ell_k$ features into a single value, and applies a tanh activation function $\tanh(x)=\frac{e^{x}-e^{-x}}{e^x+e^{-x}}$.
The module outputs are then combined using a linear classifier, described in the main text. The network architecture for the third iteration $k=3$ is shown in Fig.~\ref{fig:example_network}.
Note that in our implementation, for the first module $m^1$ we only apply a single convolution layer with linear activation, which can approximate all mappings from a binary input to a real valued output.

\section{Correlation functions in different basis}\label{app:differentbases}
Here, we discuss correlations functions of the TFIM in different measurement basis. Conventionally two helpful correlation functions in the Ising model are the two-point correlation function $C_2(l)=\langle \sigma^z_{m+l}\sigma^z_m\rangle$ and the string (parity) order correlation function $C_s(l)=\langle \prod_{m=1}^l\sigma^x_{m}\rangle$, which both are sharp indicators for the ordered (disordered) phase~\cite{PFEUTY197079}. Note that the parity correlation relies on the fact that the iTEBD algorithm prepares the true ground state, which in a finite system is the superposition of the two short-range correlated ground states.

Generalized to an arbitrary measurement basis, which we parametrize with the angle $\theta$, the two-point correlation function $C_2^\theta(l)=\langle \sigma^\theta_{m+l}\sigma^\theta_m\rangle$ and the string order correlation function $C_s^\theta(l)=\langle \prod_{m=1}^l\sigma^\theta_{m}\rangle$ are expressed as functions of the spin $\sigma^\theta_l=\cos(\theta)\sigma^z_l+\sin(\theta)\sigma^x_l$. For $\theta=0$ this corresponds to correlation functions measured in the $z$ basis while $\theta=\frac{\pi}{2}$ correspond to measurements in the $x$ basis. We also consider intermediate bases with $\theta = 0.18\pi, \pi/4$, and $0.32\pi$. 

Correlation functions like $C_2^\theta, C_s^\theta$ can be reconstructed from individual snapshots (projective measurements) by averaging over many different positions (due to translational invariance), 
and are thus accessible to the machine learning algorithm as well. 
Furthermore, in the $z$ and $x$ bases, these are sharp identifiers of the corresponding phase, and also witness the scaling with distance $l$. It is therefore likely that the algorithm learns (a variant of) these correlation functions. However, these are two of the simplest correlations functions one can consider, and most generally the network can capture any linear combination of correlation functions constructed from $\sigma^\theta$.

\begin{figure*}
    \centering
    \includegraphics[width=0.95\textwidth]{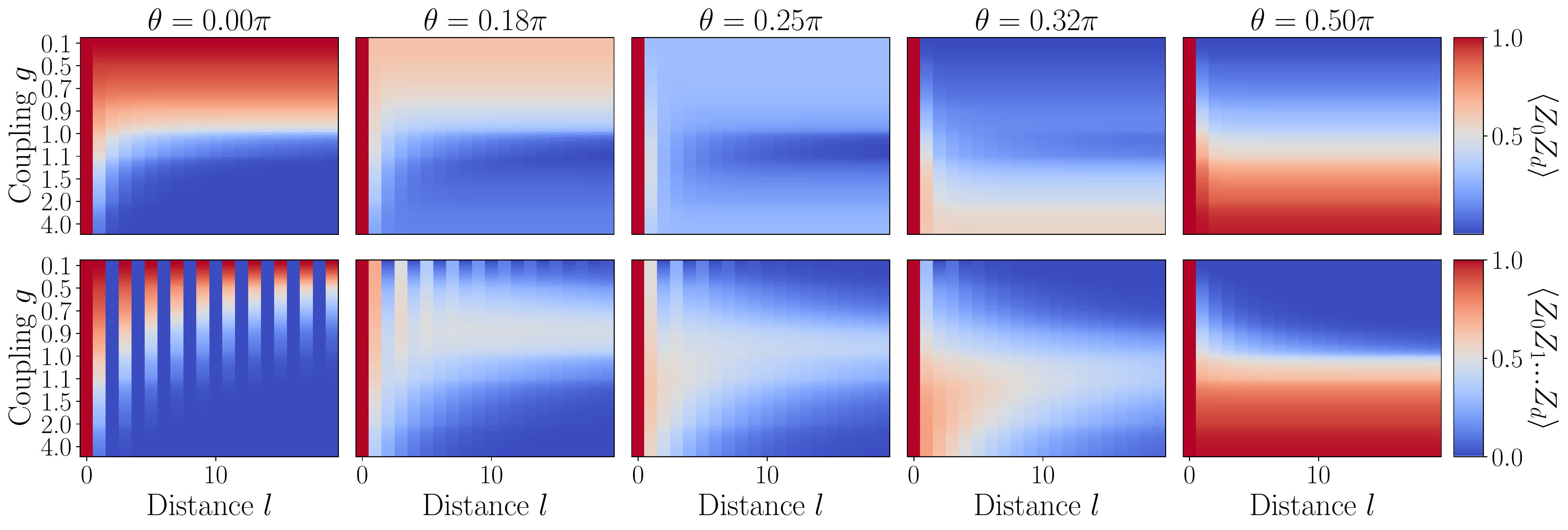}
    \caption{Two-point correlation function $C^\theta_2(l)$ (top row) and string order correlation function $C^\theta_s(l)$ (bottom row) obtained from the ground state of the TFIM in different measurement basis (parametrized by $\theta$). We see that $C_2(l)$ exhibits scaling with $l$ in the $z$ basis, but not in the $x$ basis, while the opposite is true for $C_s(l)$.
    Away from the critical point, depending on the angle $\theta$, at least one of the two correlation functions unambiguously distinguishes the two phases. An exception is the peculiar behavior at $\theta=\frac{\pi}{4}$, which we discuss in the text.}
    \label{fig:supp_correlationsA}
\end{figure*}

In Fig.~\ref{fig:supp_correlationsA}, we display the correlation functions $C^\theta_2, C^\theta_s$ for a set of angles $\theta$, obtained from the ground state for different values of $g$. We observe that in the intermediate bases, the contrast in these correlation functions on the two sides of the transition is reduced. Most strikingly at $\theta = \frac{\pi}{4}$, both correlation functions appear identical away from the critical point, and only change qualitatively close to $g=g_c$.
Thus, the two phases are not readily distinguishable by studying these two ``conventional'' correlation functions in this basis, yet the algorithm learns to distinguish them successfully.
\newline

\section{Different labellings}
\begin{figure}[h]
    \centering
    \includegraphics[width=0.48\textwidth]{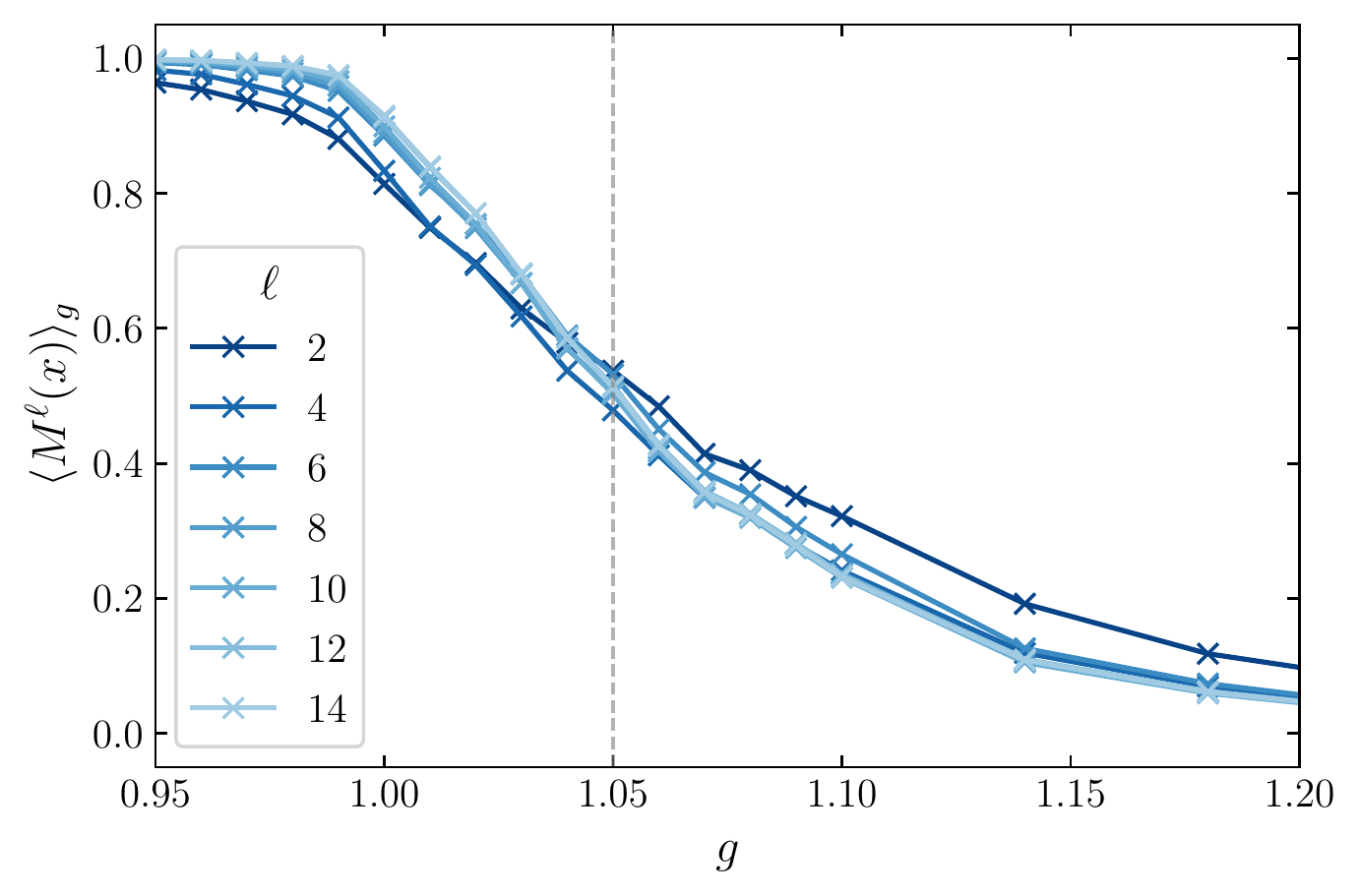}
    \caption{Classifications curves for the incorrectly chosen partition $g_{th}=1.05$. Notice that the classification improves as we increase $\ell$ for $\ell < 5$, but then saturates and stops improving. This is likely a reflection of the finite correlation length at $g_{th}=1.05$. }
    \label{fig:supp_shifted_curves}
\end{figure}

In the main text, we always partition our data into ordered (1) and disordered (0) based on knowledge of the critical point $g_c$. However, in many experimental settings, the critical point may not be known a priori. Hence, in this section, we present some results where a different threshold $g_{th}$ was used to partition the data. Specifically, we chose $g_{th}=1.05$, label snapshots from $g<g_{th}$ ordered, snapshots with $g > g_{th}$ disordered, and repeat the same training procedures. However, when we look at the classification $\langle M^\ell \rangle$ on a validation set, we see the the accuracy stops improving beyond $\ell \approx 6$, see Fig.~\ref{fig:supp_shifted_curves}. This likely reflects the finite correlation length at $g_{th}=1.05$, versus the divergent correlation at $g=1.00$, which is what enabled improvement at larger $\ell$ in main text Fig.~\ref{fig:classification_scaling}.
This also suggests that scaling of $M^\ell$ could be combined with confusion methods~\cite{van_nieuwenburg_learning_2017} to aide in identification of critical points.

\end{document}